# On the potential contribution of rooftop PV to a sustainable electricity mix: the case of Spain

Antonio Gomez-Exposito[1] (*), Angel Arcos-Vargas, and Francisco Gutierrez-Garcia

**University of Seville, Spain**

**Abstract:**
This work evaluates the potential contribution of rooftop PV to the future electricity mix. Several sustainable scenarios are considered, each comprising different shares of centralized renewables, rooftop PV and storage. For each generation scenario, the storage capacity that balances the net hourly demand is determined, and the portfolio combination that minimizes the cost of supplying electricity is obtained. The analysis is applied to mainland Spain, using public information and detailed granular models, both in time (hourly resolution) and space (municipal level). For the Spanish case, when the flexibility of hydro and biomass generation is taken into account, the least-cost portfolio involves rather modest storage capacities, in the order of daily rather than seasonal values. This shows that a sustainable, almost emissions-free electricity system for Spain is possible, at a cost that can be even lower than current wholesale market prices.

*Highlights:*
- *Overall PV rooftop potential is evaluated and is higher than expected.*
- *A bottom-up, near-zero emission electricity model is proposed at the granular (municipalities & hourly) level.*
- *Full electrification of light-duty transport is duly considered.*
- *Optimal storage needs are close to average daily demand.*
- *When foreseeable sustainable generation assets are considered, the resulting LCOE is similar to current wholesale market prices.*
- *The required distributed PV capacity is compatible with the network capacity.*

---

---

[1] (*) Camino de los descubrimientos s/n. 41092 Seville (Spain). age@us.es

# 1. Introduction.

Many countries are pledging for renewable electricity shares of 75-80% by 2030 and nearly 100% by 2050. In this context, the impact of a mostly renewable generation mix on power systems will ultimately depend on the share of centralized generation assets versus distributed ones, mainly composed of rooftop photovoltaics (PV) deployed in urban settlements.

Indeed, the cost reduction of small-scale renewable plants, energy storage systems (EES) and electric vehicles (EV), the availability of smart meters and the growing social concern about environmental issues [1], [2] have promoted a favorable state of opinion regarding self-sufficiency initiatives based on clean, behind-the-meter technologies. Germany, with over 1.5 million rooftop systems, Hawaii, and Australia constitute paradigmatic examples of this decentralized approach to electricity procurement [3] [4]. On the other hand, relevant stakeholders are rather focusing their business models on the current centralized generation paradigm, which is tantamount to replacing large thermal units by renewable plants of similar size [5]. China and India, with PV plants exceeding 1 GW [6], are the best representatives of this model [7] (Fig. 1 shows the time evolution of centralized and distributed global PV capacities).

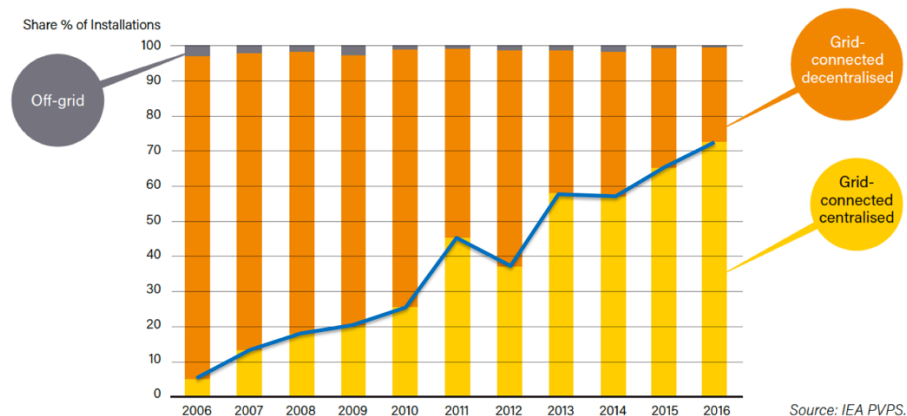

*Figure 1*: *Time evolution of centralized and distributed cumulative global PV capacities over the last decade.* The trend clearly shows a steady increase in the share of utility-scale PV plants, probably due to the widespread adoption of capacity auctions in many countries [8]

Clearly, centralized plants feature lower capital costs, but distributed facilities provide other advantages for the whole system, including their lower grid requirements and, if installed on rooftops, reduced or null additional surfaces [9] At this crucial moment, when it is still uncertain whether the ambitious commitments pledged for the COP21 Paris agreement will be achieved [10], it is of outmost importance for policy makers, investors and grid planners to have an accurate estimate of the potential contribution of urban renewable resources to a decarbonized energy mix.

**Table 1**, provides a brief account of recently published works on these issues, including their contribution, strengths, and weaknesses [11-19]. A more detailed description of those works can be found in the Supplementary Note 1.

*Table 1* . *Strengths & weaknesses of related works*

|  | **Jacobson et al. 2013, 2018, Declucci and Jacobson 2012, & Jacobson and Declucci 2013.** | **DNV, 2018** | **Ram et al., 2019** | **Shell, 2018** |
|---|---|---|---|---|
| **Scope** | North America municipalities by 2050 | Worlwide by 2050 | Worlwide by 2050 | Worlwide by 2070, UE and USA by 2050 |
| **Contribution** | Development of a roadmap to transition to a renewable scenario based on 100% of Wind, Water and Sunlight generation. | Creation of a model to estimate the energy sources in the future | Technical feasibility and socio-economic viability analysis for a transition of the global electricity system to 100% renewable generation. | Definition of a pathway for decarbonising the global economy. |
| **Strengths** | 100% of renewable energy generation | Electrification of some part of transport and building consumption. Calculation based on data from previous years adapting the estimation model to current scenario. | Hourly values of consumption and generation. Transition to 100% renewable energy production. Accurate storage system definition. | Electrification of transport by road and part of building end-use energy. Transition to a 100% renewable scenario. |
| **Weakness** | Use of yearly annual values of consumption and production. Electrification of all energy end-use and transport. | Use of yearly annual values of consumption and production. No decarbonization scope. Regionally inaccurate. | No electrification of any other sector (transportation, building, etc). Aggresive transition model. | Regionally inaccurate. Very long time-scale simulation. Use of yearly annual values for cosumption and production. |
| **Criticism** | Strong response from Trainer, T. (2012, 2013). |  |  |  |

Our work differs from previous ones in that it mainly focuses on the potential role and contribution of rooftop PV to the electricity mix, using Spain (about 30 million customers located in over 8,000 municipalities) as a case study, which can be easily replicated in other countries. Distinguishing features of our work are: 1) it builds up a national electricity model by integrating the hourly PV production and demand data at the municipal level (bottom-up approach); 2) it takes advantage of existing buildings and electrical assets, incorporating society's expectations (elimination of nuclear energy and reduction of emissions); 3) it considers several combinations of generation and energy storage capacity, identifying the one that minimizes the total cost of serving electricity nationwide; 4) the extra demand arising from the electrification of the light-duty vehicles is included.

This work considers several scenarios, some of them including the current and foreseeable portfolio of sustainable generators, and determines suitable combinations of rooftop PV and storage ensuring that the net hourly demand is fully supplied. The costs associated to each scenario are estimated, which can help policy makers in shaping the best energy regulation and incentives to achieve a sustainable electricity mix.

**2. Material and methods.**

As stated, the first goal of this work is to provide a thorough assessment of the potential contribution that rooftop PV can make to the electricity mix of a whole country in a decarbonized future, for which Spain is used as a relevant case study. The aim is to classify each conurbation as self-sufficient or not, in terms of rooftop PV production, and to estimate the storage capacity that will be required to achieve the energy balance for a given period (day, month, season), as well as the resulting costs.

The second goal is to evaluate the feasibility and cost of several brownfield sustainable scenarios, combining the foreseeable centralized generation with the rooftop PV that will be needed in each case to serve 100% of the demand. For each scenario, the storage capacity required to achieve the energy balance on an hourly basis is determined. Then, among all possible combinations of rooftop-PV and storage satisfying the expected demand, the one that minimizes the average cost of supplying electricity is obtained.

A diagram showing the input data, the main methodological steps and the scenarios considered is provided in *Figure 2*.

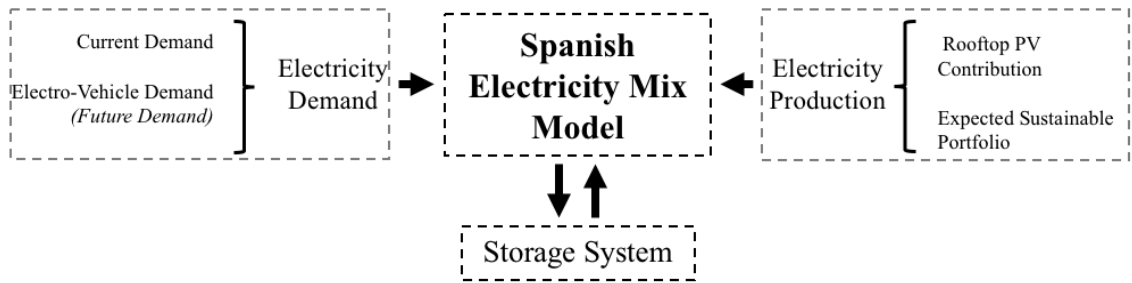

*Figure 2*: *Overview of the input data, processing blocks and results involved in the proposed assessment.* The hourly demand, including that of the EV, is to be balanced by the hourly production of the partly manageable sustainable generation system, which calls for a certain storage capacity to cover possible gaps.

Overall, four separate procedures are sequentially considered in this work, namely:

1. Determination of hourly demand profiles on a municipal basis: publicly available data bases, along with statistics regression analysis, are used to determine the annual electricity demand of each municipality, which is then split by days and hours in accordance with the typical load profiles provided by the Spanish TSO (REE) for each day of the year. Industrial loads and EV's are separately considered from the rest of consumption.

2. Determination of hourly rooftop PV production on a municipal basis: starting from the total urban surface of each municipality, the available rooftop surface is estimated. Then, depending on the solar irradiance corresponding to each place, the expected hourly PV production is calculated for each municipality.
3. Hourly electricity balance on a municipal basis: based on the hourly electricity demand and the feasible PV production for each municipality, a balance is performed to determine which urban entities, over which periods of time, are self-sufficient, and how much storage is needed. Several scenarios are considered in this assessment, with and without EV demand, each differing in the amount of battery storage assumed, from daily to seasonal horizons. The most cost-effective combination of PV and battery storage is also determined at the national level for projected unitary costs.
4. Hourly electricity balance including the existing and foreseeable renewable and cogeneration assets: unlike in the previous item, dealing with the worst-case theoretical situation in which rooftop PV is the only electricity source, the existing and foreseeable renewable power (including pumped hydro), plus the highly efficient cogeneration systems, are duly considered for the hourly balance. This reduces the need for battery storage and the associated costs.

When applied to the Spanish case, the following main hypotheses are assumed:

- Only the continental Spain, accounting for 93% of the population and 94% of the national electricity demand, will be considered. This way, the analysis is restricted to a big and rather homogeneous interconnected system, avoiding the specifics of the Canary and Balearic islands.
- In the baseline scenario (EV's excluded), the electricity demand will be assumed to be about the same as that of 2018. Indeed, the decarbonized economy of the future will be more electrified, but also more efficient. So, at this stage, it is uncertain how much, if any, the electricity demand will grow in one or two decades from current levels. This way, when matching demand with production, we will use the actual hourly profiles from 2018.
- Likewise, it will be assumed that the number of EV's remains the same as in the current ICE fleet, over 23 million cars plus about 6 million EV equivalent, corresponding to buses, motorcycles and vans in 2018.

Each major procedure is described in more detail in the following subsections.

**2.1. Determination of hourly demand profiles on a municipal basis.**

In this work, the hourly electricity demand is estimated on a granular (municipal) basis for two different scenarios: 1) base case scenario, in accordance with the actual demand (2018), implicitly assuming that the demand increase associated with other uses in the next two decades will be approximately compensated by expected efficiency gains, most notably those arising from the adoption of near-zero energy buildings, heat pumps, etc.; 2) same as base case scenario, but assuming 100% of the light-duty vehicle fleet is electrified, which is added to the current demand.

As the annual electricity demand is not available for all municipalities in Spain, it is estimated by means of an econometric model which resorts to the following data, taken from Instituto Nacional de Estadística [20]: population, annual per capita income, cadastral value (excluding industrial land), elevation above sea level (altitude) and climatic zone. The model is adjusted with complete information on electricity consumption available for the municipalities of Andalucía, Pais Vasco and Comunidad de Madrid [21-23].

To ensure consistency with the actual consumption, the estimations obtained for each municipality are scaled so that the sum extended to each region equals the actual demand provided by the System Operator [24] Once the annual demand at the municipal level is estimated, it must be prorated on an hourly basis throughout the 365 days of a year. To this end, it is assumed that the average load profiles provided by the Spanish Ministry [25] on an hourly basis are of application at the municipal level as well.

A more detailed description of this procedure is provided in Supplementary Note 2.

In addition to the base case consumption, the demand arising from the electrification of 100% of the light-duty vehicle fleet is considered in this work. For this purpose, starting with the stock of vehicles (buses, motorcycles, vans, cars, etc.) for each municipality [26] all categories are transformed into an equivalent fleet of cars in terms of annual consumption. Then, hourly demand values are derived from data provided by the System Operator [27], which regularly monitors a pool of charging points. See the Supplementary Note 3 for more details on this procedure.

## 2.2. Determination of hourly rooftop PV production on a municipal basis

The second step of the proposed methodology is aimed at estimating the maximum hourly rooftop PV production on a municipal basis, for which two procedures are combined. On the one hand, it is necessary to estimate the total available rooftop surface on every municipality; on the other, the estimation of the solar potential is performed at every location, considering the surface slope and orientation. Overall, the resulting model is similar to the one developed in [28] which performs a geospatial assessment of rooftop solar PV for the European Union, yet on a more granular basis.

The assessment of the available rooftop surface is carried out using the application ArcGIS (Geographical Information System) along with cartographical data from the national geographical database ([29], for the study case analyzed). The following hypotheses have been adopted:
- The rooftops can be divided in two different types: pitched and flat roof.
- For pitched roofs, only the four main orientations are considered (South, East, West, North), with balanced roof distribution (i.e., 25% of pitched roof on average to each orientation).
- The surface utilization factor, i.e., the fraction of rooftops that can accommodate PV panels, is taken from a previous assessment performed for the city of Seville [30].

As analyzing the roofs of each municipality individually would be very time-consuming, the urban areas are clustered according to their size and climatic zone. Then, a canonical representative for each group is studied in detail and the results are extrapolated to all members of the cluster. A sample of municipalities are manually tested *a posteriori* to check the adequacy of this approach.

The information necessary for the calculation of the hourly electricity generation, with PV panels deployed on the rooftops, comprises three factors: 1) available surface for PV installation (estimated as explained in the Supplementary Note 4), 2) the performance of the PV panels, and 3) the hourly irradiation over the respective urban surfaces. By combining the hourly irradiation for each orientation and inclination, the available surface for each type of rooftop and the size and peak power of the deployed PV panels, the energy production over the whole year is obtained. A similar approach is developed by Ordoñez et al in [31]. who roughly estimate the rooftop PV contribution in Andalusia based on monthly values for each subregion.

The hourly distribution of PV electricity production is then estimated from the data provided by the Photovoltaic Geographical Information System (PVGIS) tool developed by the European Commission [32] PVGIS gives an hourly irradiation distribution, for a typical day of every month, depending on the geographic location, PV technology and the orientation and slope of the panels.

The irradiation values for the twelve canonical days are translated into available energy per kWp through the energy conversion factor, also obtained from PVGIS tool. The hourly electricity production for the remaining days is obtained by means of an interpolation algorithm programmed in MATLAB and applied to every rooftop orientation. As the solar radiation hitting the north oriented rooftop surfaces is about a half of that received by those oriented to the south, only the pitched rooftops oriented to south, east and west, along with the flat ones, are considered when estimating the PV potential of urban rooftops (i.e., only 75% of the available pitched rooftops are actually accounted for).

Finally, by duly integrating all of the above factors, the maximum installable power and hourly electricity production throughout the year is estimated for each municipality and, as a byproduct, for every province and region of Spain. On average, each kWp of rooftop PV is estimated to yield about 1,500 kWh/year, roughly 9% less than the 1,650 kWh/year produced in 2018 by a utility-scale kWp in Spain. Further details of the procedure for estimating the surface area of rooftops, as well as the required information for the Spanish case, are provided in the Supplementary Note 4.

**2.3. Hourly electricity balance between rooftop PV and demand on a municipal basis**

Given the hourly demand for each municipality, and the hourly production of the maximum PV power installable on its rooftops, the net hourly balance is obtained simply by subtracting those two magnitudes. As will be seen below, the annual PV production exceeds the demand on an aggregated basis, which means that, with the introduction of sufficient storage capacity, capable of dealing with daily and seasonal imbalances, it is theoretically possible to design a self-sufficient electric energy system, free of emissions and requiring no additional land (in the sequel, this will be termed the *greenfield* scenario). A complete review on the storage size determination for renewable energy systems is presented in [33].

For each demand scenario (with and without EV's included), the associated costs (LCOE) are computed following the same methodology as in [30] to be compared with current wholesale or retail electricity prices. The cost parameters and data involved in the calculation of the LCOE are provided in Table 2.

*Table 2 . Costs and expected life of PV and storage equipment.*

| Technology | Unitary cost | Expected life (Year) | Efficiency (%) |
|---|---|---|---|
| PV | 1 €/Wp | 25 | - |
| Storage | 100 €/kWh | 13.7 [1] | 95 |

*Source:* [1], [2] *and own elaboration*

**2.4. Hourly electricity balance including the foreseeable sustainable assets**

Last but not least, the expected sustainable portfolio (ESP), comprising the existing and authorized renewable and cogeneration facilities, are additionally considered in the hourly electricity balance (in the sequel, this will be termed the *brownfield* scenario). The installed capacity, technology, hourly production and level of manageability (i.e., fraction of energy shiftable in time) associated with each of these assets are assumed to be the same as in 2018, which is deemed representative enough for the last few years.

The hourly electricity produced by the unmanageable technologies, as provided by the market operator (OMIE), is subtracted from the gross demand, yielding the net demand to be served by the manageable centralized assets in combination with rooftop PV and battery storage. In order to determine the least-cost capacity of rooftop PV and storage, capable of filling the gap between the ESP production and the demand on an hourly basis, a rather complex constrained optimization problem could be formulated. Alternatively, an iterative procedure exploring a sufficiently large number of PV-storage combinations (complete search estrategy) is adopted in this work.

The outer loop of this process begins with the minimum PV capacity that produces enough electricity to serve the annual demand gap (for this purpose, an average utilization factor of 17% is adopted). This is determined by subtracting the available manageable generation from the annual net demand (as defined above), and adding an estimate of the annual losses incurred by the battery storage system. Then, the inner loop determines the minimum storage capacity required to achieve a perfect hourly balance. After each inner loop iteration, the PV capacity is slightly increased, in an attempt to reduce the storage requirements, and the inner loop is run again. The outer loop stops when further increases of PV capacity are not useful to decrease the associated storage capacity.

With the rooftop PV capacity and hourly production provided by the outer loop, the inner loop then proceeds as follows:

1. Select an initial value of storage capacity. The first time, an abitrarily large value should be selected, to assure feasibility (e.g., the one obtained in the *greenfield* scenario). The next times, the value obtained in the previous execution of the inner loop can be chosen.
2. For every hour of the year:
    i. If the net demand is negative (surplus of unmanageable energy), then the excess of energy is stored in the batteries,
    ii. Otherwise (positive net demand),
        a. If the batteries have enough energy stored, they fully supply the net hourly demand. In this case, the battery charging/discharging losses are calculated and cumulated.
        b. If not, the remaining residual demand is fed from the existing manageable energy.
3. If, at the end of the year, there are hours for which not all the demand is served, then the size of the batteries is insufficient and must be increased, while if all the manageable energy is not consumed, the battery size must be reduced. Go back to step 2 until the minimum battery capacity that uses all the manageable energy and satisfies the hourly demand is obtained. This is the minimum battery size compatible with the PV capacity selected in the outer loop.
4. Compute the LCOE for the resulting PV-storage configuration.

As the results are very sensitive to the fraction of hydro generation which is assumed to be manageable, the process is repeated for varying degrees of hydro manageability.

The following hypotheses are established regarding the "expected sustainable portfolio" (ESP) of generation capacity in the brownfield scenario:

- Only the current biomass and cogeneration plants remain in the thermal generation category, along with onshore wind turbines, hydro plants, PV and concentrated solar power in the centralized renewables category. This way, the CO2 emissions of the electricity sector would come only from the approximately 5.7 GW of existing cogeneration plants, which are assumed to remain in the mix owing to their relatively higher efficiencies.

- In order to estimate the amount of additional storage capacity that will be needed to balance the demand, it is most important to discriminate which centralized plants are manageable. In this work, it is initially assumed that 85% of the hydropower production (i.e., excluding run-of-the-river power plants) and 100% of electricty from biomass facilities are manageable, but lower values will be subsequently considered. As cogeneration is under the control of the incumbent industries, no flexibility is assumed for this technology.

- As of writing, the projected PV and wind generation capacity with authorized connection points amounts to an additional increase of 86% and 904%, respectively, with respect to the power installed by the end of 2018.

**Table 3** provides the installed power, the expected capacity increase and the two components (manageable and not manageable) of the annual electricity generation corresponding to each technology. The hourly electricity produced by the unmanageable technologies in 2018, as provided by the market operator (OMIE), is assumed in this work.

*Table 3. Installed capacity and annual production of the generation portfolio assumed in the brownfield scenario.*

| Energy Source | Installed Sustainable Power in 2018 (GW) | Additional Expected Power (GW) | Total Expected Sustainable Power (GW) | Non-Manageable (TWh) | Manageable (TWh) | Total (TWh) |
|---|---|---|---|---|---|---|
| Cogeneration | 5.7 | 0.0 | 5.7 | 19.9 | 0.0 | 25.6 |
| Biomass | 0.2 | 0.0 | 0.2 | 0.0 | 0.7 | 0.7 |
| Wind | 23.5 | 20.2 | 43.7 | 80.7 | 0.0 | 80.7 |
| PV | 4.7 | 42.5 | 47.2 | 66.5 | 0.0 | 66.5 |
| Hydro | 20.7 | 0.0 | 20.7 | 4.8 | 27.2 | 32.0 |
| Thermo Solar | 3.9 | 0.0 | 3.9 | 3.9 | 0.0 | 3.9 |
| Total | 58.7 | 62.7 | 121.4 | 181.4 | 27.9 | 209.3 |

*Source:* [24] *and own elaboration*

### 3. Results.

#### 3.1. Rooftop PV potential for self-consumption at the municipal level

The main outcome of the procedure described in Section 2.1, a color contour map of the mainland Spain showing the annual electricity demand at the municipal level, is represented in Figure 3. Figure 4 gives an idea of how the electricity consumption is distributed among the whole set of municipalities. Note that the 666 largest municipalities (8% of the total) consume 80% of the electricity.

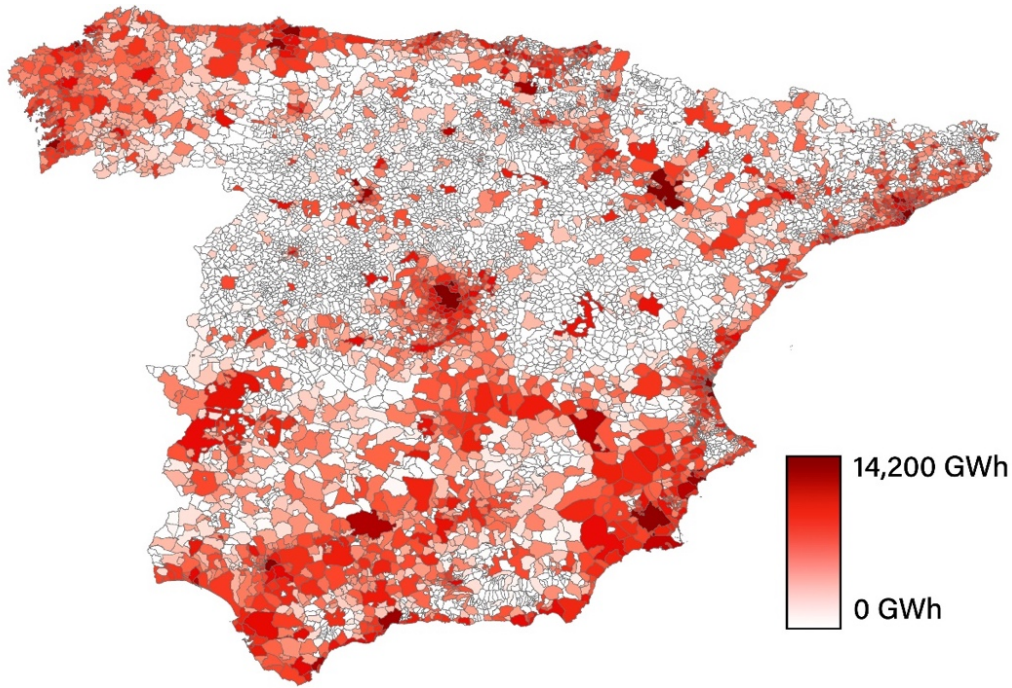

*Figure 3 :* *Annual electricity demand at the municipal level (higher demands in red).*

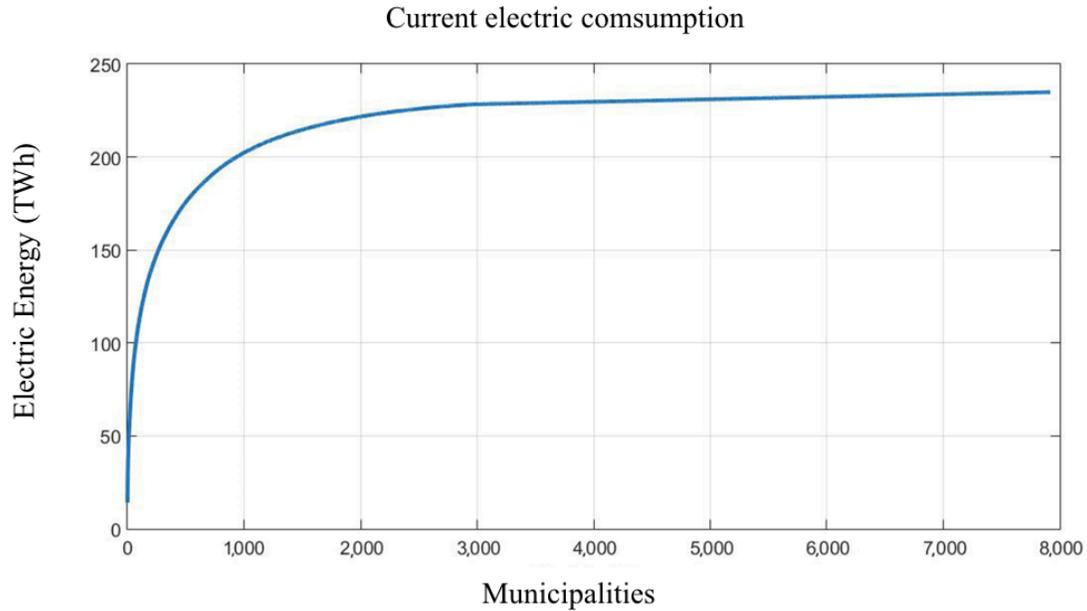

*Figure 4 :* *Lorenz curve showing the distribution of electricity consumption among municipalities.*

The rooftop PV production for each municipality, calculated in accordance to the methodology presented in Section 2.2, can be represented likewise. *Figure 5* shows the potential annual electricity production over the Spanish mainland territory for each municipality with more than 1,000 inhabitants.

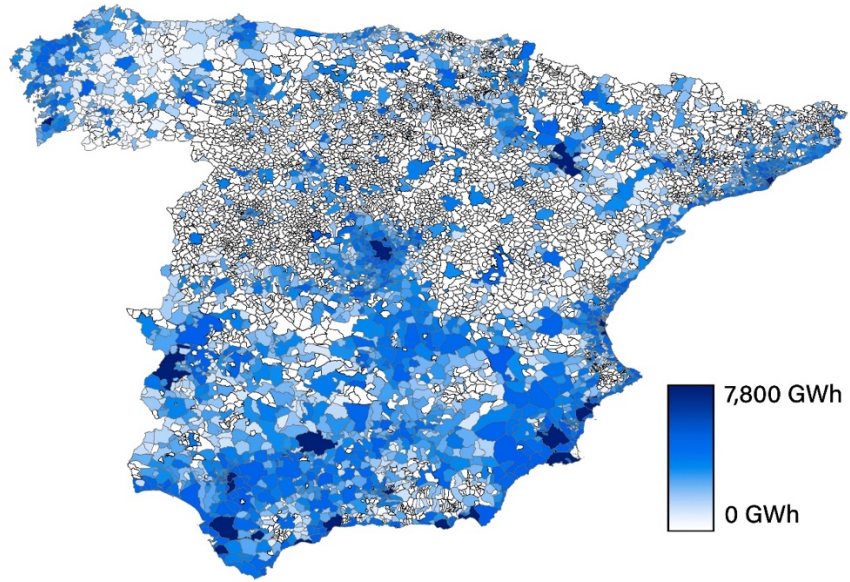

*Figure 5: Potential annual rooftoop PV production on a municipal basis.* The figure shows the maximum annual PV electricity production over the Spanish mainland territory for each municipality with more than 1,000 inhabitants [33], [26]. The most productive municipalities are obviously the big cities, as they span larger areas, but also those located in the southern regions, where the solar radiation is higher. Due to its size, Madrid would be the main producer of Spain, with a maximum estimated production of nearly 7,800 GWh.

*Table 4* provides the maximum rooftop PV power that can be installed and the annual electricity production classified by provinces, as well as the aggregated national values (234 GWp and 291 TWh, respectively). The losses due to shadow effects are considered as 10%.

*Table 4.* Maximum rooftop PV power and annual production by mainland Spanish provinces.

| Provinces | Installed PV power (MWp) | Annual production (GWh) | Equivalent hours | Provinces | Installed PV power (MWp) | Annual production (GWh) | Equivalent hour |
|---|---|---|---|---|---|---|---|
| Álava | 994 | 984 | 1,101 | La Rioja | 1,229 | 1,300 | 1,175 |
| Albacete | 2,901 | 3,673 | 1,407 | León | 4,256 | 5,158 | 1,346 |
| Alicante | 10,718 | 14,070 | 1,459 | Lérida | 2,689 | 3,194 | 1,320 |
| Almería | 5,309 | 7,623 | 1,595 | Lugo | 1,760 | 1,790 | 1,130 |
| Asturias | 2,959 | 2,834 | 1,064 | Málaga | 7,844 | 10,700 | 1,516 |
| Ávila | 1,221 | 1,467 | 1,334 | Madrid | 18,420 | 24,129 | 1,456 |
| Badajoz | 5,565 | 7,154 | 1,428 | Murcia | 10,894 | 14,219 | 1,450 |
| Barcelona | 21,609 | 27,038 | 1,390 | Navarra | 7,642 | 8,077 | 1,174 |
| Burgos | 2,118 | 2,317 | 1,216 | Orense | 2,060 | 2,283 | 1,231 |
| Cáceres | 2,835 | 3,591 | 1,407 | Palencia | 1,102 | 1,283 | 1,293 |
| Cádiz | 6,135 | 8,541 | 1,547 | Pontevedra | 5,471 | 6,007 | 1,220 |
| Cantabria | 3,768 | 3,352 | 989 | Salamanca | 1,769 | 2,172 | 1,364 |
| Castellón | 3,552 | 4,509 | 1,411 | Segovia | 1,492 | 1,746 | 1,300 |
| Ciudad Real | 4,365 | 5,574 | 1,419 | Sevilla | 11,040 | 15,123 | 1,522 |
| Córdoba | 4,740 | 6,267 | 1,469 | Soria | 705 | 849 | 1,337 |
| Cuenca | 1,789 | 2,166 | 1,345 | Tarragona | 6,139 | 7,591 | 1,374 |
| Gerona | 6,324 | 7,449 | 1,309 | Teruel | 808 | 1,015 | 1,396 |
| Granada | 6,640 | 8,936 | 1,495 | Toledo | 7,821 | 9,766 | 1,387 |
| Guadalajara | 2,067 | 2,530 | 1,360 | Valencia | 13,264 | 17,307 | 1,450 |
| Guipúzcoa | 1,845 | 1,689 | 1,018 | Valladolid | 2,615 | 3,123 | 1,327 |
| Huelva | 3,816 | 5,414 | 1,576 | Vizcaya | 3,779 | 3,373 | 992 |
| Huesca | 1,305 | 1,624 | 1,382 | Zamora | 857 | 1,047 | 1,357 |
| Jaén | 4,640 | 6,083 | 1,457 | Zaragoza | 3,837 | 4,689 | 1,358 |
| La Coruña | 9,573 | 10,090 | 1,171 | | | | |
| | | | | **Total** | **234,282** | **290,917** | **1,242** |

*Figure 6* shows the resulting annual electricity balance (surplus in blue, deficit in red) for each municipality. Note that all the ingredients necessary to perform such a theoretical assessment (demand and rooftop PV production on a municipal and hourly basis) have been laid out above.

Given that the total annual production exceeds the base case demand, it would be theoretically possible to design an electrical supply system solely based on rooftop PV installations, that uses the current distribution and transmission grids to match the geographical unbalances, and that resorts to not yet existing storage facilities to shift energy in time as needed (see more details below).

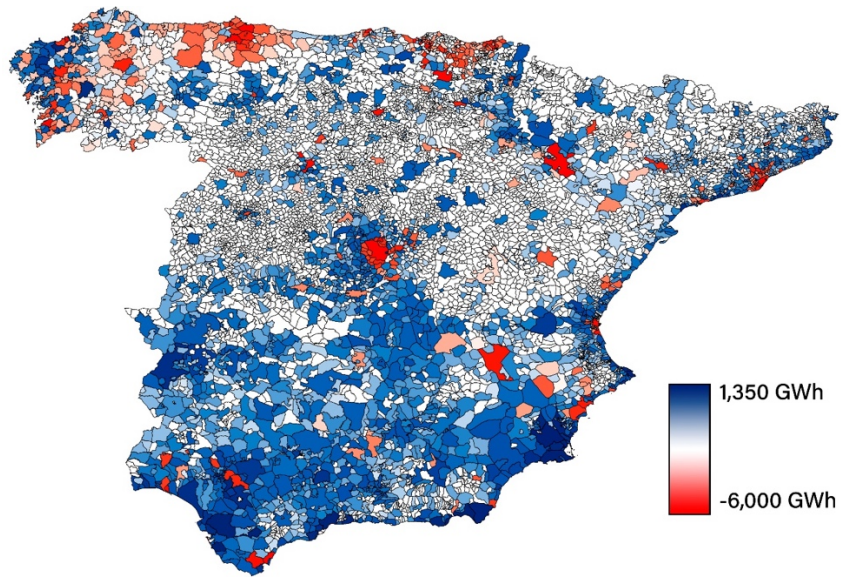

*Figure 6: Annual electricity unbalance on a municipal basis for the greenfield theoretical scenario.* The figure shows, for every municipality, the difference between the annual demand and its maximum rooftop PV annual production. Those with an annual energy deficit (large cities, industrial sites and some northern municipalities with low irradiation) are shown in red, while those with an energy surplus are presented in blue. More than 95% of the analyzed municipalities have a potential for rooftop PV production exceeding their annual demand.

The annual net balance does no give a clue of what happens on a temporal (daily, monthly or seasonal) basis, which is a valuable byproduct of the study undertaken in this work. *Figure 7* offers further insights into how the energy unbalances take place, both by provinces and months. It turns out that 11 provinces (including Madrid and Barcelona) lack enough rooftop surface to be self-sufficient in terms of PV electricity production. At the national level, the country would not be either self-sufficient over the quarter November-January, whereas Seville (the province with highest annual surplus) would be self-sufficient all the year round and Asturias (the one with highest deficit) would be always deficient if PV was deployed only on rooftop surfaces.

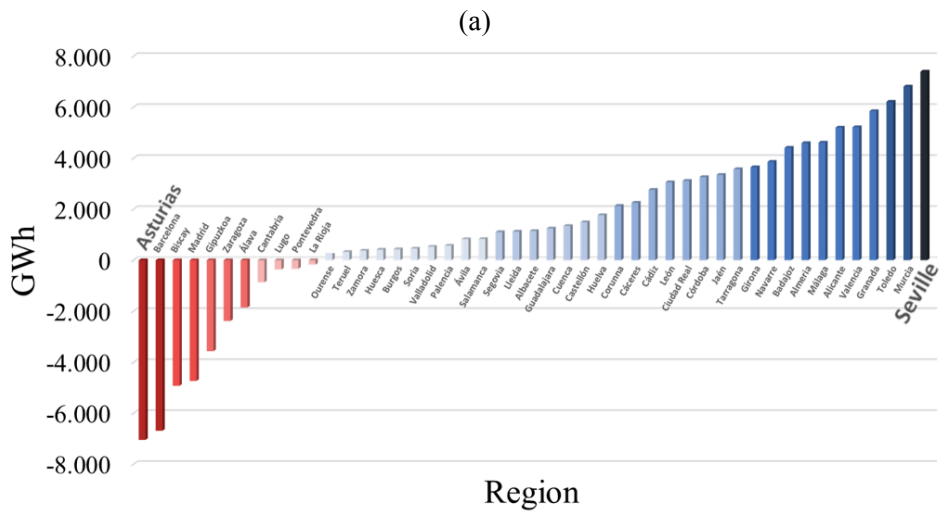

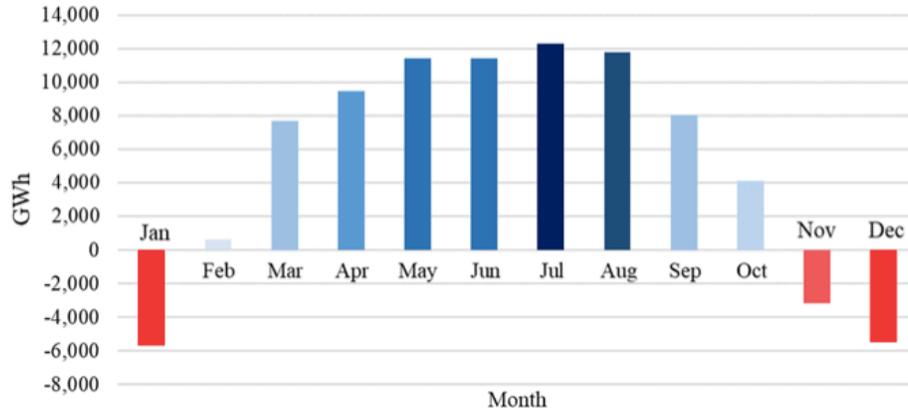

(c)

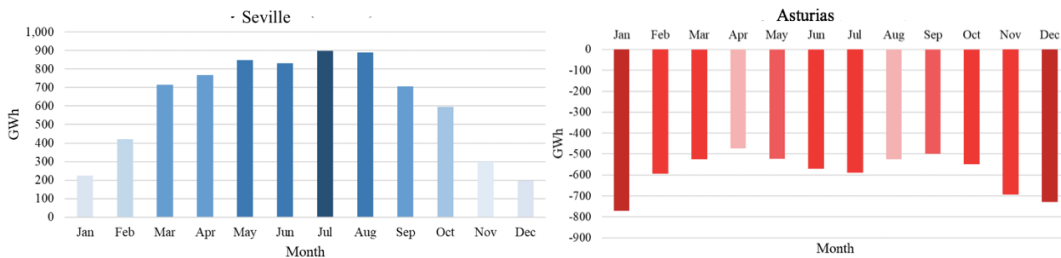

***Figure 7:*** *Annual electricity balance by provinces for the base case (a), monthly electricity balance for the total mainland Spain (b), and for the two provinces with the greatest unbalance, Seville (surplus) and Asturias (deficit), (c).* The provincial results (a) have been obtained by aggregating the municipal results presented in Figure 6. Note that, despite the fact that the annual production of the maximum installable rooftop PV is higher than the overall national demand, during the winter months this is not the case (b).

In a future scenario with all vehicles electrified (trucks excluded), we should add to the equation the 60.8 TWh of electricity that would yearly consume the equivalent fleet of cars (nearly 30 million equivalent cars) [24] [25]. Even in this case, there is sufficient rooftop surface area to accomplish the annual energy balance (295 TWh of total demand), for which a small fraction of north-faced roofs should be used. In the sequel, all scenarios will include the EV demand, unless otherwise noticed.

**3.2. Greenfield scenarios: rooftop PV plus battery storage only**

The future scenarios analyzed in this paper are not aimed to reflect a specific year, but rather refer to a hypothetical medium-term horizon in which the huge potential of both renewable sources and EV's has materialized to a large extent. This will likely happen at some time in the decade from 2030 to 2040, depending mainly on yet unveiled technological advances and how ambitious the implemented policies are.

To begin with, we consider an idealized or "thought" scenario intended to answer, among others, the following major questions:
    1) Would PV panels, massively deployed on the rooftop surfaces of today's urban settlements, produce sufficient electricity to totally satisfy the annual demand of a whole country?
    2) If so, when and where will energy unbalances (surplus or deficit) arise, and how much energy storage will be strictly needed to restore the required balance on specified time horizons (daily and seasonal)?

3) What would the electricity cost be in such idealized scenario, compared with the average cost of today's wholesale and retail electricity markets?

By applying the procedures described in the Methods section to a case study (in this work the whole mainland Spain), the right answer can be given to the questions posed above. The goal is to provide a thorough, granular and accurate assessment of the potential contribution the distributed rooftop PV can make to the electricity mix in a decarbonized future. Not only is a system-wide perspective provided, but the energy balance is also carried out by hours for each municipality. This way, for a given period (month, season, year) each conurbation can be classified as self-sufficient or not in terms of rooftop PV production, and the associated storage capacity and cost can be quantified.

As shown above, overall a maximum of 234 GWp of PV panels can be accommodated on an estimated available surface of 1,134 km$^2$ of rooftops (this excludes north-faced roofs), representing about 46% of the total urban surface and just 0.22% of the national surface. Such installed capacity can deliver annually around 290 TWh of electricity (1,240 equivalent hours on average), theoretically sufficient to supply the 235 TWh consumed by mainland Spain in 2018 (REE, 2019).

As evidenced by *Figure 7*, though, noticeable seasonal unbalances arise, which may differ significantly among provinces or regions. If no other source of firm power is available, then some energy storage will be required to shift PV power from day to night and, depending on the location, from summer to winter. In this work, several storage scenarios are considered, ranging from the most exigent case in which all excess of solar energy in summer must be shifted to the winter, to the case in which there is no storage at all, including in between the more realistic one in which the storage devices are dimensioned just to trade energy on a daily basis. A total of forty different storage scenarios have been analyzed and the results are summarized in *Figure 8*.

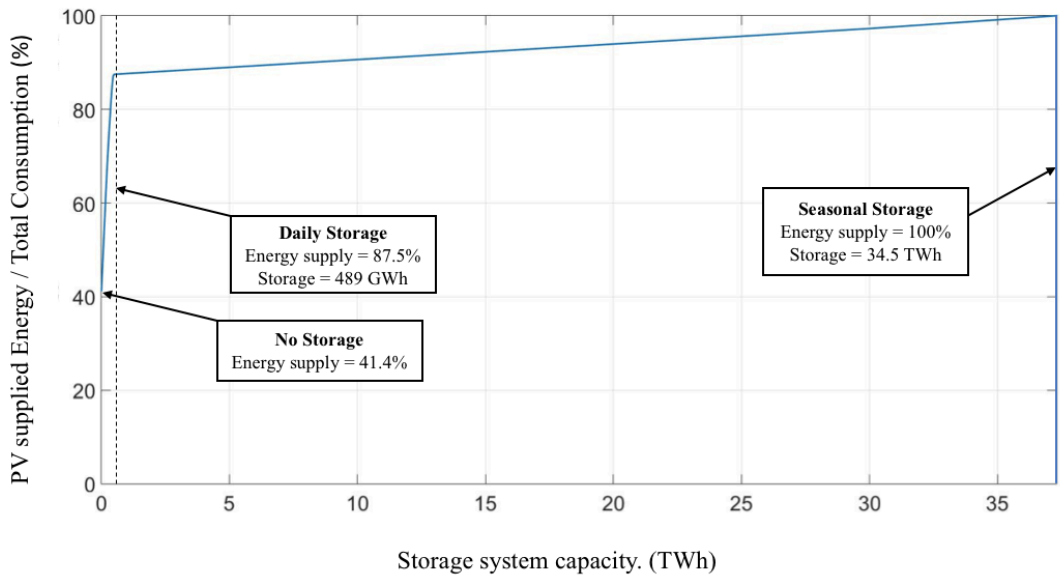

**Figure 8:** *Demand coverage vs. installed storage capacity (EV demand included).* The percentage of annual electricity supplied by the rooftop PV system is shown for increasing storage capacities. Note the strongly non-linear evolution of the demand coverage as the storage capacity increases, with a very high slope until the daily storage needs are satisfied, and then a "saturation" region (small slope) for the remaining scenarios, until 100% seasonal coverage is achieved.

In absence of storage, the coverage of the demand barely exceeds 41% (50% if EV's were excluded[2]). Note that the storage capacity required to cover 100% of the demand in this theoretical greenfield scenario (34.5 TWh) would be infeasible or disproportionately expensive with current technology. However, by ensuring sufficient daily storage (489 GWh) nearly 90% of the demand could be served.

The costs (LCOE) associated to all scenarios are computed, assuming Li-ion batteries are used for storage (*Figure 9* represents the interval around the least-cost storage capacity). It is worth noting that, while the optimal storage capacity (435 GWh) is insufficient to ensure the daily balance in the worst case, both costs are relatively close (between 50 and 52 €/MWh, of the same order of magnitude as today's wholesale European markets). Therefore, the daily storage capacity can be considered also sufficiently optimal for practical purposes, which is good news.

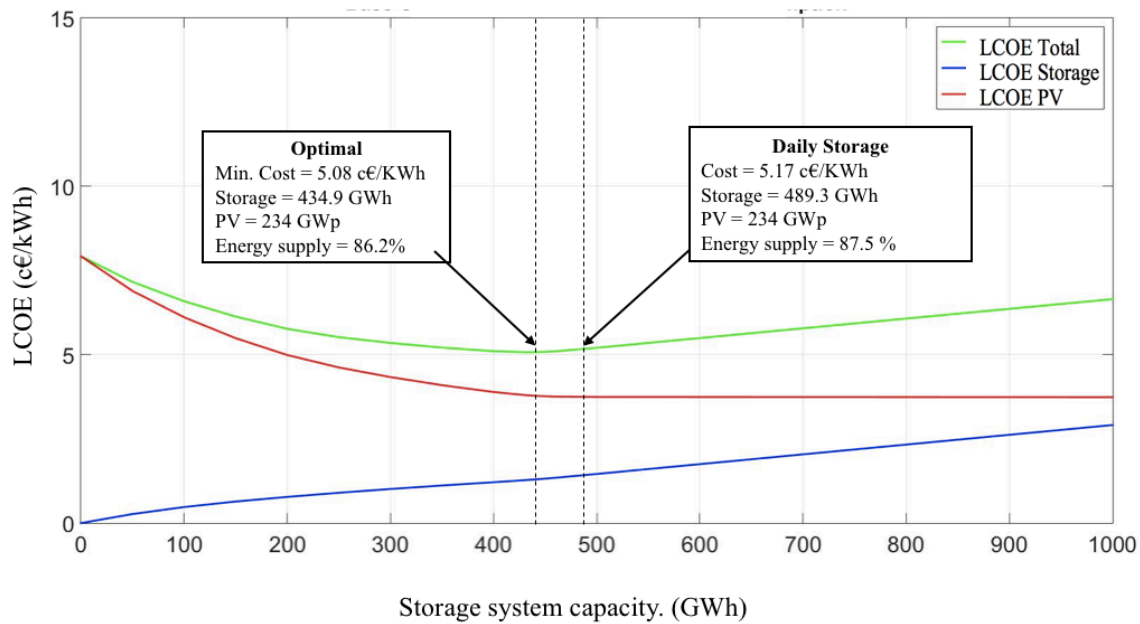

*Figure 9:* *LCOE estimation vs. battery capacity storage.* Only the area around the minimum LCOE value has been represented, as the costs associated with cases warranting 100% of demand coverage (seasonal storage) are extremely high (around 900 €/MWh).

### 3.3. Brownfield scenarios: inclusion of the foreseeable sustainable assets

The *greenfield* case constitutes just a reference scenario providing local and global upper limits to the potential share of distributed PV generation, as well as the storage capacity required to achieve daily or seasonal balances. However, more credible *brownfield* scenarios should duly consider existing and planned sustainable generation plants if a realistic assessment is to be performed of the feasibility of the resulting mix and the required investments.

In this regard, in addition to the existing renewable assets, a long-term future scenario should also consider the wind and PV capacity which has already been granted a connection point to the grid. As shown in the former section, the foreseeable capacity and annual production of such "expected sustainable portfolio" (ESP), are currently much higher than those projected by the Spanish government in the recently issued

---

[2] The results of the base case analysis, excluding the demand of EVs, are included in the Supplementary Note 5.

Integrated National Plan of Energy and Climate (PNIEC) [34]. Unlike the PNIEC, this work excludes any thermal generation, except for the biomass facilities and 5.7 GW of existing cogeneration plants, the latter being kept in the mix owing to their relatively higher efficiencies. In order to estimate the additional storage capacity that will be needed to balance the hourly demand, it is most important to discriminate which centralized plants are manageable. Initially, this work assumes that 85% of the hydropower production (run-of-the-river power plants excluded) and 100% of biomass electricty are manageable.

In total, 209.3 TWh are expected to be produced by the centralized fleet of ESP generators, of which only 27.9 TWh are initially assumed to be manageable. In absence of any other resource, this is clearly insufficient to serve the foreseeable demand (295 TWh, EV including). Moreover, a detailed hourly analysis shows that, despite the manageable fraction of energy being used in an optimal fashion, there are intervals of time for which the net demand is negative, which calls for unmanageable generation curtailment. The data collected in **Table 5** shows that nearly 100 TWh (32%) remain userved in this case, whereas 5.5% of the ESP electricity should be curtailed

*Table 5. Annual demand, production and balance for the brownfield scenario in absence of rooftop PV and battery storage*

| Current scenario | | Energy (TWh) | Total Energy (TWh) |
|---|---|---|---|
| Demand | Current Demand | 234.9 | |
| | EV Demand | 60.8 | 295.7 |
| Production | Non-Managebale energy | 27.9 | |
| | Managebale energy | 181.4 | 209.3 |
| Balance | Total demand | | 295.7 |
| | Used energy | | 197.8 (94,5 %) |
| | Unserved demand | | -97.9 |

We assume that the additional electricity required to close the gap between the ESP production and the expected demand (i.e., the residual demand) is produced exclusively by rooftop PV. For the sake of sizing the associated storage system, it is also assumed that the manageable energy can be arbitrarily shifted in time, as long as the required power is lower than the rated power of the involved generators (the details of how the hourly balance is carried out in this case can be found in Section 2.4). Unlike in the greenfield scenario, comprising only distributed assets, a single national balance is performed in this case, as there is no reasonable way of splitting or decomposing the production of centralized assets on a municipal basis. This implicitly assumes that the transmission system is designed so that the electricity can flow across the Spanish geography as required. **Table 6** provides the breakdown of battery storage and PV costs in the optimal case.

*Table 6. Rooftop and storage capacity, production and costs in the least-cost brownfield scenario.*

| Assessment | Technology | Unite of measurement | Value |
|---|---|---|---|
| New systems | Installed rooftop PV power | [GWp] | 75.0 |
| | Rooftop PV usuful energy | [TWh] | 93.0 |
| | Storage capacity | [GWh] | 298.0 |
| Investment | PV | [M€] | 75,000 |
| | Storage | [M€] | 29,800 |
| | Total | [M€] | 104,800 |
| Depreciation | PV (25 years depreciation) | [M€/Year] | 3,000 |
| | Storage (13.7 years depreciation) | [M€/Year] | 2,175 |
| | Total | [M€/Year] | 5,175 |
| LCOE | - | [c€/kWh] | 5.58 |

*Figure 10* represents the isoquant curve of all rooftop PV-storage combinations that can feed the total net demand, clearly showing two asymptotic values. Assuming unlimited rooftop surface for PV production, the minimum energy storage capacity is 190 GWh, while the minimum PV power capable of serving the remaining net demand, including the losses due to charge-discharge battery cycles, is 66 GWp. The figure also shows the resulting LCOEs for all isoquant combinations. Note that the minimum cost (55.8 €/MWh) is achieved for a rooftop PV capacity of 75 GWp and a storage system of 298 GWh. Such cost is slightly lower than the average wholesale market price in 2018 in Spain (56.4 €/MWh). Moreover, the resulting *mix* is virtually emission-free and, being distributed to a larger extent, would reduce the overall system losses and network requirements.

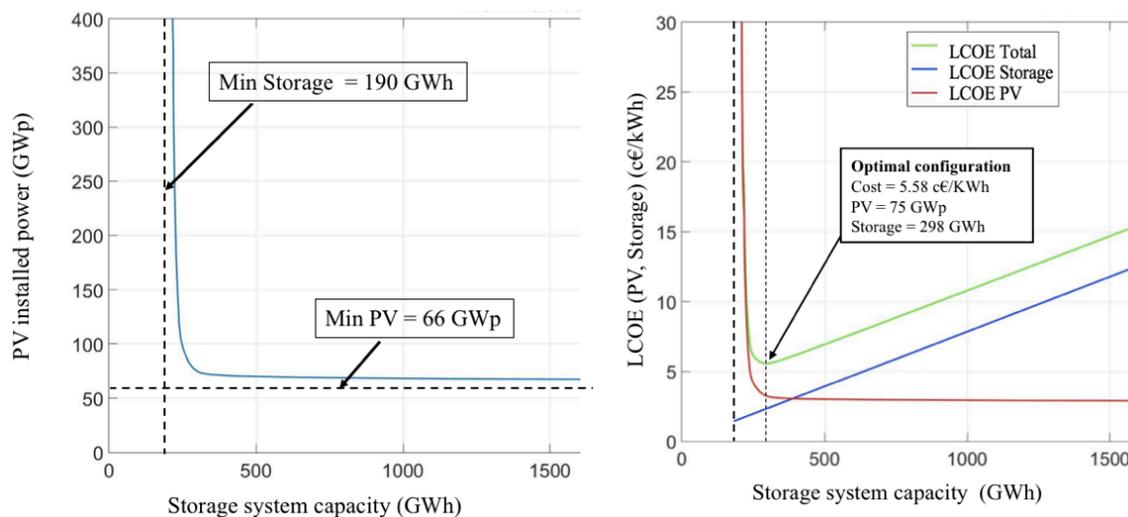

*Figure 10: Isoquant curves showing all possible combinations of rooftop PV and storage capacity serving 100% of the net residual demand in the brownfield scenario (left) and related LCOEs (right).* This scenario assumes that 85% of hydro is manageable. The combination leading to minimum LCOE is marked up on the left diagram.

One of the parameters most affecting the results presented in *Figure 10* is the fraction of hydro electricity which is manageable. In order to assess the sensitivity of the results to this parameter, the analysis is repeated for decreasing levels of hydro manageability (from 40 to 85%). *Figure 11* represents the LCOEs for the additional isoquant rooftop PV-storage curves, showing how the minimun cost evolves for each level of hydro manageability. As expected, the lower the proportion of manageable hydro, the higher the cost, as more rooftop PV and storage must be added to the system.

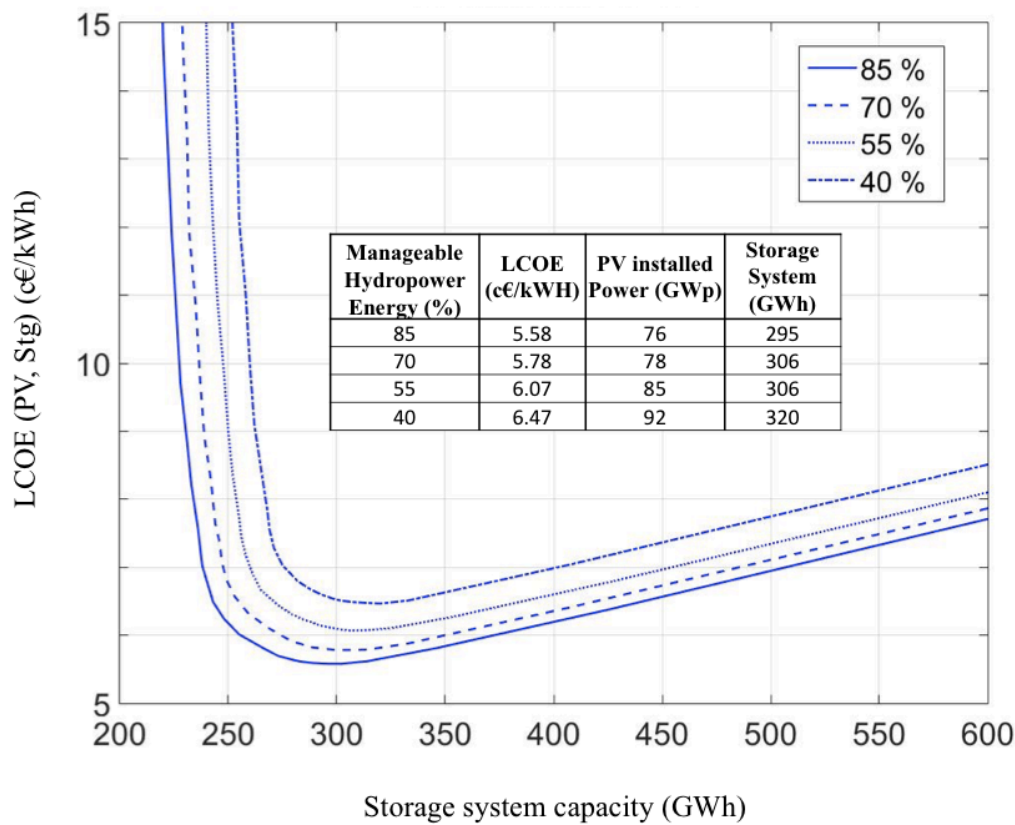

*Figure 11: Minimun LCOE for each level of hydro power manageability.* In addition to the total LCOE of Fig. 10, this figure shows the resulting costs for lower levels of hydro manageability. The least-cost combination of PV and storage is marked up for each case, and the associated figures are provided in the embedded table, in order to show the cost sensitivity to the share of manageable generation.

**4. Discussion**

**Table 7** summarizes the main results for each of the five scenarios considered in the paper, including the optimal combination of rooftop PV and storage, the percentage of demand coverage, curtailed energy (which could be used for other purposes, such as water desalination or $H_2$ production), as well as the associated costs. For the computation of the averge LCOE, it has been assumed that the cost of the electricity provided by the ESP assets corresponds to the average cost of the whosale market in 2018.

**Table 7.** *Least-cost combination of rooftop PV-battery storage capacities under different sustainable scenarios.*

| Scenario | | Annual Load | | Manageable ESP. | | Unmanageable. ESP. | | PV | | Battery Storage | | Served Energy | Curtailed Energy | LCOE (New system) | Whosale Market 2018 | Average Energy Cost |
|---|---|---|---|---|---|---|---|---|---|---|---|---|---|---|---|---|
| | | Energy | Peak | Energy | Rated | Energy | Rated | Energy | Peak | Energy | Peak | | | | | |
| | | TWh | GW | TWh | GW | TWh | GW | TWh | GWp | GWh | GW | % | TWh | €/MWh | €/MWh | €/MWh |
| Greenfield | Base case | 235 | 41 | - | - | - | - | 291 | 234 | 294 | 60 | 91% | 80 | 55.2 | 56.4 | 55.3 |
| | Base case + EV | 296 | 55 | - | - | - | - | 291 | 234 | 435 | 94 | 88% | 31 | 50.8 | 56.4 | 51.5 |
| Brownfield (EV included) | ESP only (Hydro is 85% manageable) | 296 | 55 | 28 | 18 | 181 | 104 | - | - | - | - | 67% | 12 | 55.8 | 56.4 | 56.0 |
| | ESP + PV + Storage (Hydro is 85% manageable) | 296 | 55 | 28 | 18 | 181 | 104 | 103 | 75 | 289 | 61 | 100% | 16 | 55.8 | 56.4 | 55.8 |
| | ESP + PV + Storage (Hydro is 40% manageable) | 296 | 55 | 13 | 8 | 196 | 114 | 126 | 92 | 320 | 66 | 100% | 64 | 64.7 | 56.4 | 64.7 |

Regarding the greenfield scenarios, the following conclusions are worth stressing:

- In Spain, the available urban surface (excluding north-faced rooftops) can theoretically accommodate enough PV panels to deliver 100% of the annual electricity consumed in 2018. However, this would require unacceptably expensive amounts of seasonal battery storage. The least-cost rooftop PV-storage combination (234 GWp of PV and 294 GWh of storage) can supply 91% of the current demand at an estimated cost of 55 €/MWh.

- When 30 million equivalent vehicles are assumed to be electrified, the least-cost rooftop PV-storage combination involves 55% more storage (455 GWh) but the LCOE is lower (51 €/MWh). Should the energy stored by EV batteries be partially used as an additional grid resource (V2G scheme), the storage capacity could be drastically reduced (this possibility is not explored in this work).

- In both cases, a significant fraction of the rooftop PV production must be curtailed or used for other purposes (86 TWh in the base case and 30 TWh when the EV is considered).

When the sustainable centralized assets (ESP) are included in the balance (brownfield scenarios), the following conclusions are drawn:

- As expected, the ESP assets alone would be insufficient to satisfy the demand. Assuming 85% of the hydro production is flexible enough, still nearly 32% of the demand cannot be served.

- The least-cost rooftop PV-storage combination capable of serving, along with the ESP centralized fleet, 100% of the demand at an average cost of 56 €/MWh, comprises just 75 GWp of PV and 298 GWh of storage.

- Should the rate of hydro manageability be reduced from 85 to 40%, the least-cost capacity of rooftop PV and battery storage would increase to 92 GWp and 320 GWh, respectively, and the average cost would reach 60 €/MWh.

- Interestingly enough, in all cases the optimal combination of PV and battery storage leads to storage capacities which are in the order of the energy consumed on a daily basis. Therefore, expensive seasonal storage is not needed at all, as long as a minimum amount of manageable capacity is available. In fact, the average costs are similar to those of the Iberian wholesale market in 2018.

- The resulting generation and storage portfolio has enough firm capacity (power) to deal with the respective peak demand, even if the non-manageable ESP is fully ignored.

- The distributed PV facilities offered by the brownfield model are below the suggested one in (35), below which there would be no congestion problems on transmission and distribution systems.

- From the figures collected in **Table 7**, it turns out that the storage should be dimensioned on average for less than 5 hours of rated power, of the same order as the latest generation of stationary batteries currently being deployed worldwide.

## 5. Conclusion

This paper has first shown that the Spanish urban surfaces can accommodate enough rooftop PV capacity to potentially feed the current electricity demand plus the consumption of the light-duty vehicle fleet, should it be electrified. Very costly seasonal storage would be needed to serve 100% of the demand in this theoretical greenfield scenario. However, if approximately 10% of the demand could be served by other flexible means, then the remaining 90% could be satisfied solely by PV plus a reasonable daily storage capacity. The adopted methodology is based on public data, with hourly and municipal resolution and can be applied to any other country.

Then, other more realistic scenarios, additionally considering the expected sustainable portfolio of mostly centralized generation assets, are also analyzed. For each scenario, the storage capacity that minimizes the average cost of supplying electricity is obtained. The most relevant conclusion is that the least-cost combination of generation assets (centralized renewables, rooftop PV and storage) that satisfies the foreseeable demand (EVs included) leads to daily rather than seasonal storage capacities, with associated LCOEs of the same order as the average cost of the Iberian wholesale market.

In a nutshell, this work has proven that a sustainable, emissions-free electricity system is possible, for a big country such as Spain, by suitable and affordable combinations of centralized and distributed generation, the latter one providing up to nearly 45% of the demand, along with relatively modest amounts of storage capacity, mainly dimensioned for daily cycles.

## List of References

# Supplementary material.

# On the potential contribution of rooftop PV to a sustainable electricity mix: the case of Spain

Antonio Gomez-Exposito (*), Angel Arcos-Vargas, and Francisco Gutierrez-Garcia

University of Seville, Spain

Supplementary Note 1. Recently published studies related to the scope of this work.
Supplementary Note 2. Determination of hourly demand profiles on a municipal basis.
Supplementary Note 3. Inclusion of the light-duty EV demand.
Supplementary Note 4. Determination of rooftop surface and PV potential production.
Supplementary Note 5. Greenfield scenario: Demand coverage & LCOE.

**Supplementary Note 1. Recently published studies related to the scope of this work.**

As the search for a new energy model with low costs and emissions is the "Holy Grail" of our days, there is a plethora of research and prospective efforts focused on developing suitable frameworks and pathways, as well as defining the directions and challenges that regulators, stakeholders and researchers must face.

Among the published works, we have highlighted in this note those by Jacobson et al. (2013, 2018), Det Norske Veritashas (DNV, 2018), Lappeenranta University of Technology (LTU), developed by Ram et al. (2019), and Shell (2018), as the most comprehensive and credible ones.

The study coordinated by Jacobson carries out an assessment of the electrification of all sectors in 53 North American cities, using wind, water and sunlight (WWS) as the only sources of energy. Although it is a complete and rigorous work, in our opinion some aspects are not faced in a realistic manner. Perhaps the most questionable one is the fact that all energy uses are assumed to be electrifiable, which may not be the case of air and long-distance naval transportation and certain high-temperature industrial processes. On the other hand, it is based on average annual values, fully ignoring that the electricity demand has a large seasonal and hourly variability, which may impose severe limitations and congestion problems on transmission and distribution systems, as analysed for instance by Tevar et al. (2018). Finally, a global bottom-up vision is missing, as they limit themselves to the separate analysis of a sample of cities, pretty much in line with the model proposed by Arcos-Vargas et al. (2019) for the city of Seville, where more exhaustive results are provided. Jacobson's contribution, although valuable, has attracted many critics by the scientific community, among which that of Trainer (2012), widely debated in academic circles, can be pointed out.

Another worthwhile contribution is the model proposed by DNV, more realistically considering the baseline energy mix, including non-renewables. It electrifies neither all transport nor the consumption and provides a projection of future demand based on increases in population, efficiency and productivity. Yet, as in the previous case, the analysis is restricted to annual values of energy consumption, not duly considering the seasonal and daily variations, while the underlying network is assumed of infinite capacity for the transport of energy. A major drawback of working with just annual balances lies in the fact that the calculated storage is over dimensioned, as happens also in the work by Gomez-Exposito *et al.* (2018).

A more sophisticated assessment is performed by LTU (Ram et al. 2019), which does take into account the seasonal and hourly nature of the demand, as well as the network limitations. This way, more moderate storage values are obtained, although the quick transformation assumed for the next decades seems to be unrealistic in view of the current panorama.

Finally, unlike the previous assessments, which are targeted to 2040 or 2050, the Shell model (Shell, 2018) designs a roadmap to 2070, in which the population and energy growths provided respectively by the United Nations and the IEA are considered. This study, of a more descriptive nature than the previous ones, carries out a prognosis over too long a period, lacking details of transmission or storage technologies, which leaves the model weakly defined. Furthermore, as in most previous cases, it is restricted to annual consumption values, and divides the world into 10 regions, which precludes the use of a bottom-up approach.

The methodology adopted in this work is similar to that proposed in Arcos-Vargas et al. (2018) for the city of Seville, where it has been shown that the potential of rooftop PV for latitudes like that of Spain is huge, largely exceeding 50% of the overall energy needs.

**Supplementary Note 2. Determination of hourly demand profiles on a municipal basis.**

As stated in Section 2, the goal is to decompose as accurately as possible the annual demand of the continental Spain (235 TWh in 2018), among the 7,918 municipalities covering the total territory. Then, for each municipality, the annual demand has to be distributed throughout the year on an hourly basis.

The histogram in *Supplementary Figure 2.1* clearly shows that the Spanish population is highly dispersed. With an average of almost 6,000 people per municipality, the distribution of conurbations ranges from the 57 cities with more than 100,000 people to the nearly 5,000 rural entities inhabited by less than 1,000 people (INE, 2019)

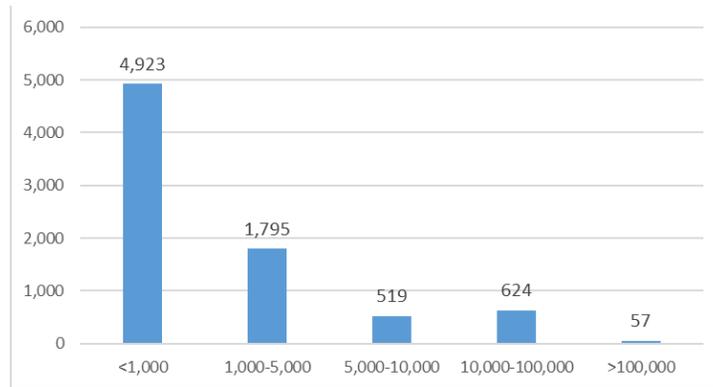

***Supplementary Figure 2.1***: *Histogram of Spanish municipalities according to the number of inhabitants. Source: INE (2019) and own elaboration*

**Supplementary Table 2.1** provides useful data regarding the relative contribution of the very small villages in terms of electric consumption, people and surface. As shown, this group represents over 40% of the total municipalities, nearly half of the surface area, about 4% of the population and less than 3% of electricity consumption. Therefore, for the sake of simplicity and manageability of data, this large set of smallest villages is clustered into 47 equivalent (virtual) entities, each corresponding to and assumed to be located in the center of gravity of an administrative province. This way, while still being able to allocate and discriminate among 97% of the population, the number of municipalities to be managed reduces from nearly 8,000 to only 3,042 (2,995 actual + 47 virtual).

***Supplementary Table 2.1*** *Relative contribution of municipalities with less than 1,000 people to the grand total.*

| Cities/Towns | Electric Consumption (GWh) | | Inhabitants (Thousands) | | Surface (Km²) | |
|---|---|---|---|---|---|---|
| | Absolute Value | % | Absolute Value | % | Extension | % |
| Total | 234,884 | - | 43,151 | - | 493,518 | - |
| <1000 hab | 6,517 | 2.77% | 1,621 | 3.76% | 204,711 | 41.48% |

*Source: Instituto Nacional de Estadística (2019) and own elaboration.*

Fortunately, three out of the 15 continental regions, Madrid (center), Andalusia (south) and Basque Country (north), altogether representing 34% of the total demand, have created specialized bodies (e.g. energy agencies) which keep regularly an account of how, when and where electricity is consumed (Junta de Andalucía. 2019, Comunidad Autonoma del Pais Vasco. 2019, and Comunidad Autonoma de Madrid.

2019). Therefore, for each of the 861 municipalities belonging to those regions, the annual electricity demand is known by categories (residential, industrial, etc.). For the remaining 12 regions (2,180 municipalities), only the annual aggregated demand is provided by REE, as shown in **Supplementary Table 2.2.**

***Supplementary Table 2.2.*** *Annual electricity consumption for the 15 mainland Spanish regions in 2018.*

| Region | GWh | Region | GWh |
|---|---|---|---|
| Cataluña | 44,569 | Asturias | 9,953 |
| Andalucía | 35,457 | Aragón | 9,843 |
| Madrid | 28,925 | Región de Murcia | 7,424 |
| Comunidad Valenciana | 24,367 | Navarra | 4,578 |
| Galicia | 18,570 | Extremadura | 4,441 |
| Pais Vasco | 16,501 | Cantabria | 4,326 |
| Castilla y León | 12,939 | La Rioja | 1,581 |
| Castilla La Mancha | 11,410 | **Total** | **234,884** |

*Source: REE (2019)*

In order to obtain an estimate of the electricity demands of all the continental Spanish municipalities, an econometric method has been developed, based on demand data from the 861 municipalities of those three regions, by using of five explanatory variables:

- population,
- annual per capita income,
- cadastral value (excluding industrial land),
- elevation above sea level (altitude) and
- climatic zone (peninsular Spain is divided into seven climatic zones).

Then, the coefficients of both regressions are applied to the municipalities of the 12 remaining regions for which no detailed consumption data exists, and the resulting electricity demands are properly scaled so that the sum of municipal demands for each region agrees with the aggregated values provided by REE (2019), this yields a regression fit of $R^2$=0.97, which is considered very satisfactory. ***Supplementary Figure 2.2*** shows the actual demand versus the value estimated by our model for the whole set of known municipalities.

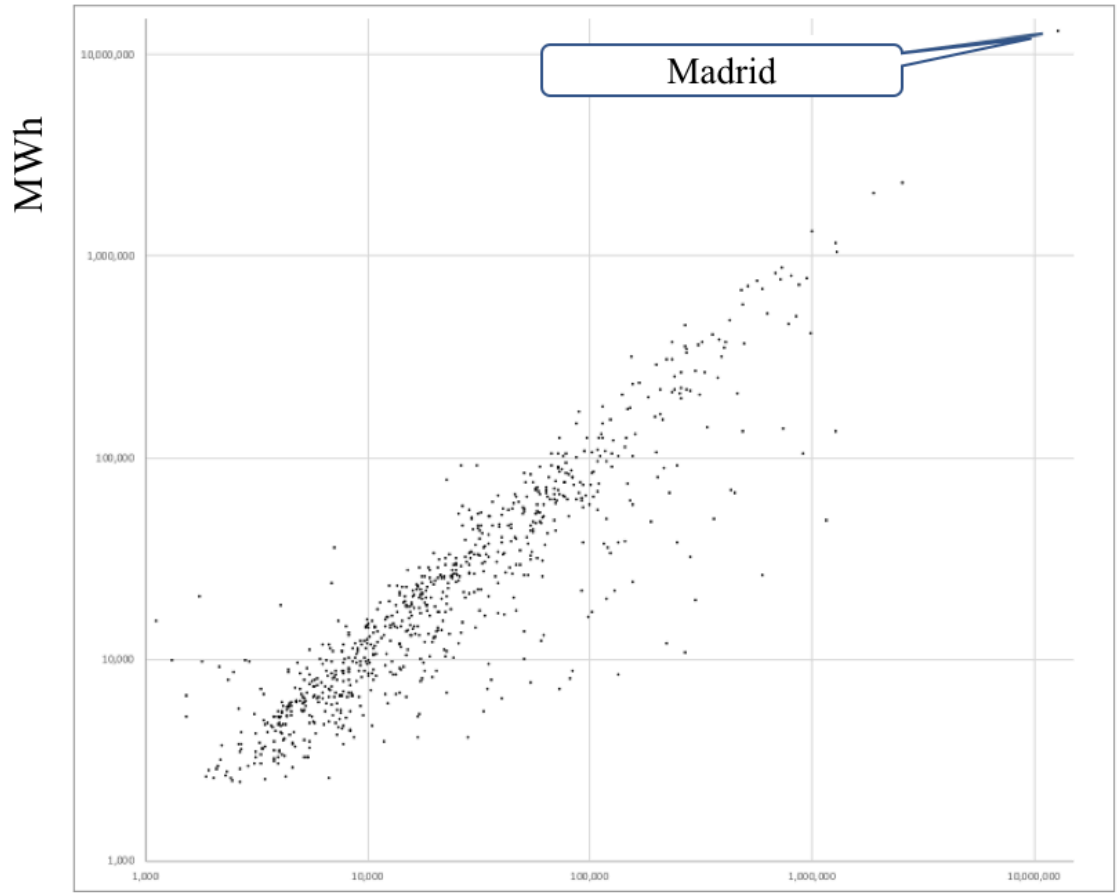

**Supplementary Figure 2.2:** *Expected (vertical axis) vs. actual (horizontal axis) electricity consumption of the municipalities located in Madrid, Andalusia and Basque Country.*

**Supplementary Note 3. Inclusion of the light-duty EV demand.**

At the moment, it is uncertain how many and which types of vehicles will be electrified in two decades. On the one hand, the population in Spain is not expected to grow significantly (in fact, it has reached a plateau phase). On the other hand, future mobility trends seem to favor car sharing and automated transportation, most likely leading to a fleet reduction. For these reasons, as stated above, it will be assumed that 100% of the existing ground transportation fleet will be electrified. Trucks will be excluded from the analysis, as their daily mileage range is well beyond what current battery technologies can provide. Indeed, unless a technological breakthrough in batteries (e.g., in flow batteries) or wireless *en route* charging takes place, the most promising scheme to decarbonize trucks, ships and aircrafts seems to be through synthetic fuel (gas or liquid), including hydrogen-based fuel cells.

The number and types of vehicles registered and paying taxes at each municipality is provided by the Dirección General de Tráfico (Ministerio del Interior, 2019). **Supplementary Table 3.1** summarizes that information for the 15 continental regions in Spain, along with the expected electricity consumption if the entire fleet was electrified.

*Supplementary Table 3.1.*
Number of vehicles, by regions and categories, and expected electricity consumption.

| CCAA | Cars | Vans | Buses | Motorbikes | Equivalent Vehicles | Consumption (MWh) |
|---|---|---|---|---|---|---|
| Andalucía | 4,080,704 | 412,476 | 9,294 | 655,261 | 5,502,931 | 11,212,221 |
| Aragón | 608,550 | 70,529 | 1,510 | 76,624 | 839,855 | 1,711,205 |
| Asturias | 516,400 | 46,942 | 1,455 | 57,057 | 687,698 | 1,401,184 |
| Cantabria | 305,540 | 25,170 | 633 | 39,225 | 394,443 | 803,677 |
| Castilla La Mancha | 1,073,946 | 129,610 | 2,318 | 112,011 | 1,475,577 | 3,006,489 |
| Castilla y León | 1,308,808 | 135,096 | 3,286 | 131,829 | 1,761,840 | 3,589,748 |
| Cataluña | 3,527,529 | 387,006 | 9,361 | 834,704 | 4,940,936 | 10,067,158 |
| Comunidad Valenciana | 2,567,237 | 198,275 | 4,599 | 386,780 | 3,273,957 | 6,670,687 |
| Extremadura | 589,861 | 71,163 | 1,364 | 55,018 | 811,916 | 1,654,280 |
| Galicia | 1,538,995 | 116,219 | 4,742 | 158,605 | 2,007,546 | 4,090,374 |
| La Rioja | 147,630 | 17,458 | 263 | 16,633 | 200,398 | 408,311 |
| Madrid | 3,759,902 | 394,281 | 11,040 | 364,437 | 5,127,741 | 10,447,772 |
| Navarra | 326,338 | 39,010 | 849 | 35,386 | 452,961 | 922,909 |
| Pais Vasco | 1,002,752 | 83,124 | 3,475 | 131,951 | 1,345,558 | 2,741,575 |
| Región de Murcia | 759,531 | 66,871 | 1,882 | 111,279 | 1,004,569 | 2,046,809 |
| **Total** | **22,113,723** | **2,193,230** | **56,071** | **3,166,800** | **29,827,926** | **60,774,400** |

*Source: Ministerio del Interior (2019) and own elaboration.*

**Supplementary Table 3.1** provides also the number of "equivalent" cars, i.e., the number of average cars that would consume yearly the same total energy. The resulting conversion factors, taking into account their average yearly mileage and their specific consumption, are given in **Supplementary Table 3.2**. This way, in terms of electricity consumption, a bus is for instance equivalent to 35 cars on a yearly average basis, leading to a total of nearly 30 million equivalent cars.

*Supplementary Table 3.2.*
Number of equivalent cars for several vehicle categories.

| Vehicle Type | Average Annual Travelled Distance (km) | Specific Consumption (Wh/km) | Average Anual Consumption (MWh) | Equivalent Vehicle |
|---|---|---|---|---|
| Cars | 12,500 | 163 | 2 | 1.00 |
| Vans | 19,500 | 235 | 5 | 2.25 |
| Buses | 55,000 | 1,269 | 70 | 34.24 |
| Motorbikes | 7,700 | 68 | 1 | 0.26 |
| Motorcycles | 11,000 | 32 | 0 | 0.17 |

*Source: Ministerio del Interior (2019) and own elaboration.*

Overall, in a future scenario with all vehicles electrified (trucks excluded), the equivalent fleet of cars shown in **Supplementary Table 3.1** would annually add about 60 TWh of electricity consumption to the 235 TWh of the base case (25.5% increase).

Currently, there are about 5,000 public charging points in Spain, many of them monitored by REE from a national control center (CECOVEL) specially developed for this purpose (REE, 2019b). This center provides hourly information, on an individual and aggregated basis, of daily charging profiles, which are also used for load forecasting. In this work, the aggregated charging profile provided by REE, properly scaled with the respective number of equivalent cars, is used to determine the additional electricity that will be hourly consumed by the fleet of EV's at the municipal level. This is based on the assumption that the charging patterns of vans, buses and motorbikes will be similar to that of cars, which is the best we can do in absence of more detailed data.

**Supplementary Note 4. Determination of rooftop surface and PV potential production.**

The methodology to analyse the potential electricity production of rooftop PV is based on the combination of two assessments. On the one hand, it is necessary to estimate the total available rooftop surface on every municipality; on the other, the estimation of the solar potential is performed at every location, considering the surface slope and orientation.

In turn, the surface evaluation comprises two stages. First of all, the cartographical data from the national geographical database is processed by ArcGIS software, providing the horizontal projection of the buildings for every municipality of Spain. Secondly, the total area is split into pitched and flat roof surfaces for every municipality, which, considering the surface utilization factor and the even distribution of pitched roofs (one fourth of the surface to each orientation) yield the final available surface for each roof type and orientation.

The analysis of the available rooftop surface is carried out using the application ArcGIS (Geographical Information System) and the public geographical (Ministerio de Fomento, 2019, for the study case analyzed).

Total building surface calculation.

The procedure for estimating the total surface occupied by buildings is based on the calculation, with ArcGis, of the area represented in the cartographic data. In **Supplementary Figure 4.*1***, for instance, a comparison between ArcGIS representation and a satellite image is shown. As can be seen, the geometry defined in the cartographic data is very similar to the real projection of the buildings. However, outside of big cities and towns, the data also contain many polygons representing other large surfaces, such as lakes or crop fields. In order to select only the actual building surfaces, an algorithm has been developed retaining only small enough polygons, which are coincident with buildings (houses and blocks).

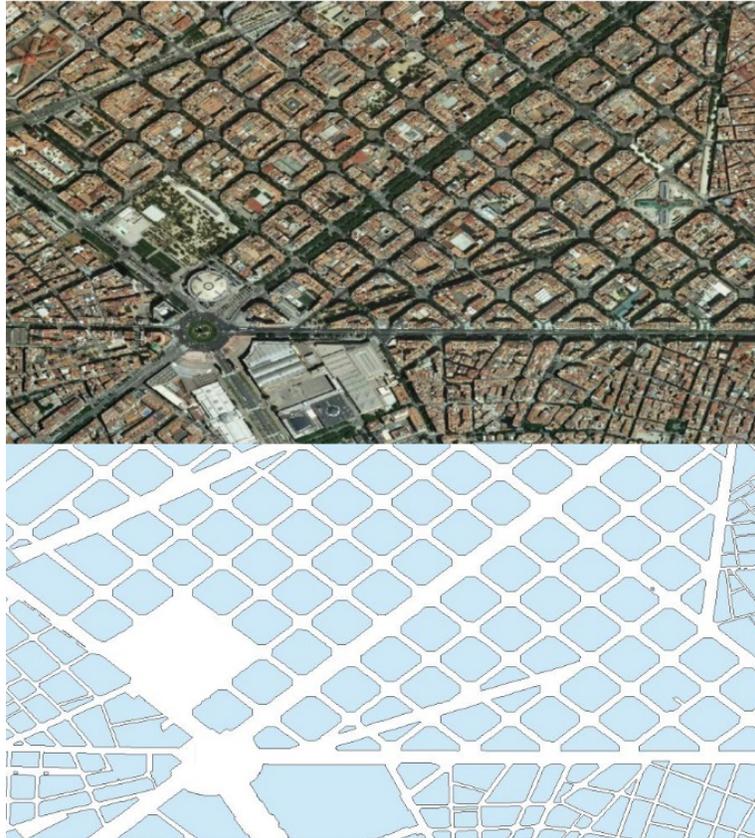

***Supplementary Figure 4.1:*** Comparison between Satellite image and ArcGIS geometry.

In order to check the reliability and accuracy of the algorithm, the results for the city of Seville are compared with those provided by the exhaustive district-level analysis carried out in Arcos et al., (2019). The building surface obtained with the new simplified algorithm is 14,952,586 m2 whereas the surface calculated in Arcos et al. (2019) is 15,065,602 m2 (out of 75 km2 of total urban surface). As the relative error is less than 1%, the proposed methodology, based on ArcGIS and the national geographical database, is considered reliable enough.

Available rooftop surface.

Once the total rooftop surface is estimated, the next step is aimed at establishing the fraction of this surface that is available for the installation of PV panels. Also, a classification of the rooftops according to their orientation and slope is needed in order to duly assess the expected electricity production.

As the scope of this study embraces over 3,000 municipalities, it is infeasible to analyze the configuration of the rooftops for all urban areas on an individual basis. Regarding the classification in flat or pitched roofs, the proposed methodology first analyzes a training sample of municipalities sharing common features and, afterwards, the results are extrapolated to the remaining population. Several test samples are then randomly selected and analyzed in detailed to compare the results and validate the estimated distribution and type of rooftops.

For this purpose, two of the parameters used in the econometric model for the demand calculation have been considered for each municipality, namely the population, divided into three ranges, [1,000-10,000), [10,000-100,000) and [>100,000] inhabitants, and the climatic zone. The latter factor significantly affects the architecture, including the shape and slope of the rooftop, with noticeable differences between the Northern and Southern regions. Using a sample of 56 municipalities, the percentage between pitched and flat rooftop is determined for each population range and climatic zone, and then applied to the remaining urban areas within each category. A few huge cities have been analyzed individually, as they cannot be assimilated to any of the categories considered. **Supplementary Table 4.1** summarizes the distribution of rooftops as a function of the previous parameters, also pointing out some outliers (Andalucía and Valencia) for the climatic regions IV and V.

*Supplementary Table 4.1:* Distribution of rooftop inclination.

| Climatic Zone | <10,000 hab | | 10,000 - 100,000 hab | | > 100,000 hab | |
|---|---|---|---|---|---|---|
| | Flat | Pithced | Flat | Pitched | Flat | Pitched |
| Zone I | 10% | 90% | 10% | 90.00% | 20% | 90% |
| Zone II | 10% | 90% | 20% | 80.00% | 40% | 60% |
| Zone III | 20% | 80% | 30% | 70.00% | 30% | 70% |
| Zone IV | 10% | 90% | 30% (40 % Valencia) | 70% (60 % Valencia) | 50% | 50% |
| Zone V | 20% | 80% | 40% (80 % Andalucía) | 60% (20 % Andalucía) | 70% | 30% |

The results shown in *Supplementary Table 4.1* have been validated by choosing a random municipality for each inhabitant range and climatic.

Once the rooftop type (flat or pitched) fraction is determined for each urban surface, the utilization factor and the orientation (just for the pitched case) have to be estimated. For the pitched rooftop orientation, considering that buildings are aligned with streets and roads, which in turn are not expected to follow any biased pattern, it is assumed that any of the four main orientations has similar chances to be found (25% of surface allocated to each one). Regarding the percentage of available area, out of the total rooftop surface, a 68% surface utilization factor is adopted. Such a value was obtained in Arcos-Vargas et al. (2019) for the city of Seville, considering permanent elements, shadows and security distances to the edges. As the city of Seville, with over 700,000 people, comprises a representative enough sample of different districts, ranging from detached houses, with one or two floors, to tall condominium blocks, this value is considered representative enough for practical purposes.

**Supplementary Table 4.2** provides the total area, the urban area and the available area of the rooftop, with the pitch considered.

*Supplementary Table 4.2.*
Avalaible area: Total, urban and rooftop.

| Spain Mainland | Area | |
|---|---|---|
| | Km2 | % |
| Total area | 493,518 | 100.00 |
| Total urban area | 2,440 | 0.49 |
| Available rooftop area | 1,354 | 0.27 |
| Plain | 475 | 0.10 |
| South oriented | 220 | 0.04 |
| North oriented | 220 | 0.04 |
| West & East oriented | 440 | 0.09 |

Estimation of PV electricity generation

The information necessary for the calculation of the hourly electricity generation, with PV panels deployed on the rooftops, comprises three factors: 1) available surface for PV installation (estimated in the previous section), 2) the performance of the PV panels, and 3) the hourly irradiation over the respective urban surfaces. By combining the hourly irradiation for each orientation and inclination, the available surface for each type of rooftop and the size and peak power of the deployed PV panels, the energy production over the whole year is obtained. Some simplifying assumptions are made:

- The solar irradiation over the capital city of each of the 15 mainland regions is adopted for each municipality of the respective region.
- The energy production of 1 kWp of panels oriented south is the same as that of panels installed on flat rooftop with optimum inclination. For other orientations, an inclination of 30o is assumed.
- The layout and specifications of the PV panels are the same as in Arcos-Vargas *et al.* (2019) for the city of Seville, namely the area occupied on average by 1 kWp power is 5.8 m2, whereas the shadow losses due to neighbor panels are 10%. This layout can be reasonably extrapolated to all mainland Spain, as the maximum difference among the optimal slope angles for all regions is 4$^o$.

The hourly distribution of PV electricity production is estimated from the data provided by the Photovoltaic Geographical Information System (PVGIS) tool (European Comission, 2012), developed by the European Commission. PVGIS gives an hourly irradiation distribution, for a typical day of every month, depending on the geographic location, PV technology and the orientation and slope of the panels.

The values of irradiation for the twelve canonical days are translated into available energy per kWp through the energy conversion factor, also obtained from PVGIS tool. The hourly electricity production for the remaining days is obtained by means of an interpolation algorithm programmed in MATLAB and applied to every rooftop orientation. As the solar radiation hitting the north oriented rooftop surfaces is about a half of that received by those oriented to the south, only the pitched rooftops oriented to south, east and west, along with the flat ones, are considered when estimating the PV potential of urban rooftops (i.e., only 75% of the available pitched rooftops are actually accounted for).

Finally, by duly integrating all of the above factors, the maximum installable power and hourly electricity production throughout the year is estimated for each municipality and, as a byproduct, for every province and region of Spain. On average, each kWp of rooftop PV is estimated to yield about 1,500 kWh/year, roughly 9% less than the 1,650 kWh/year produced in 2018 by a utility-scale kWp in Spain.

**Figure 5** in the main text shows the potential annual electricity production over the Spanish mainland territory for each municipality with more than 1,000 inhabitants.

**Supplementary Note 5. Greenfield scenario: Demand coverage & LCOE.**

The following figures are the counterparts of Figures 8 and 9 in the main text, when the demand of the equivalent fleet of EVs is ignored.

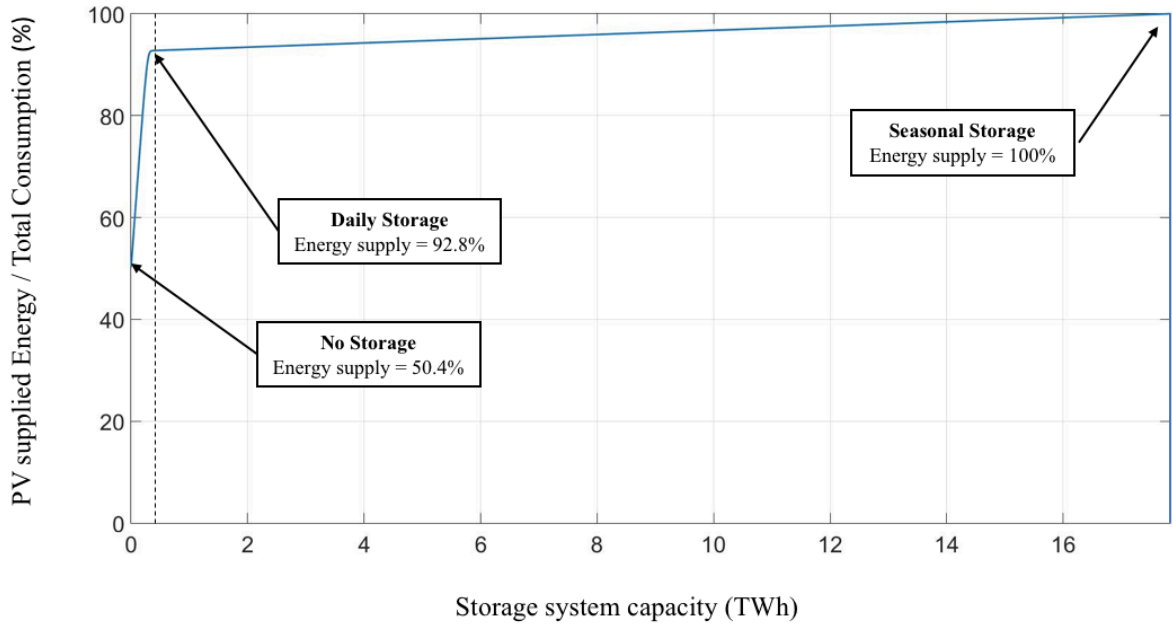

***Supplementary figure 5.1:*** *Demand coverage vs. installed storage capacity (base case).* The percentage of annual electricity supplied by the rooftop PV system is shown for increasing storage capacities.

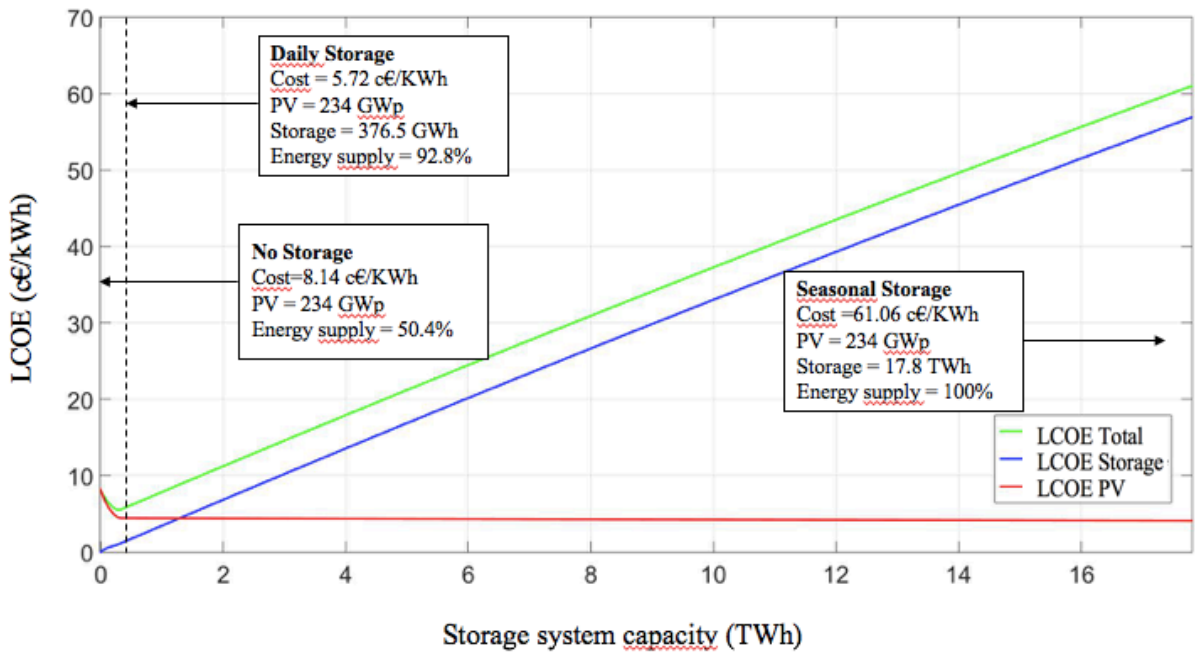

***Supplementary figure 5.2:*** *LCOE estimation vs. battery capacity storage (base case). The procedure for the cost estimation and the hypotheses considered are described in the Method 3. The figure shows the global picture from no storage to seasonal.*

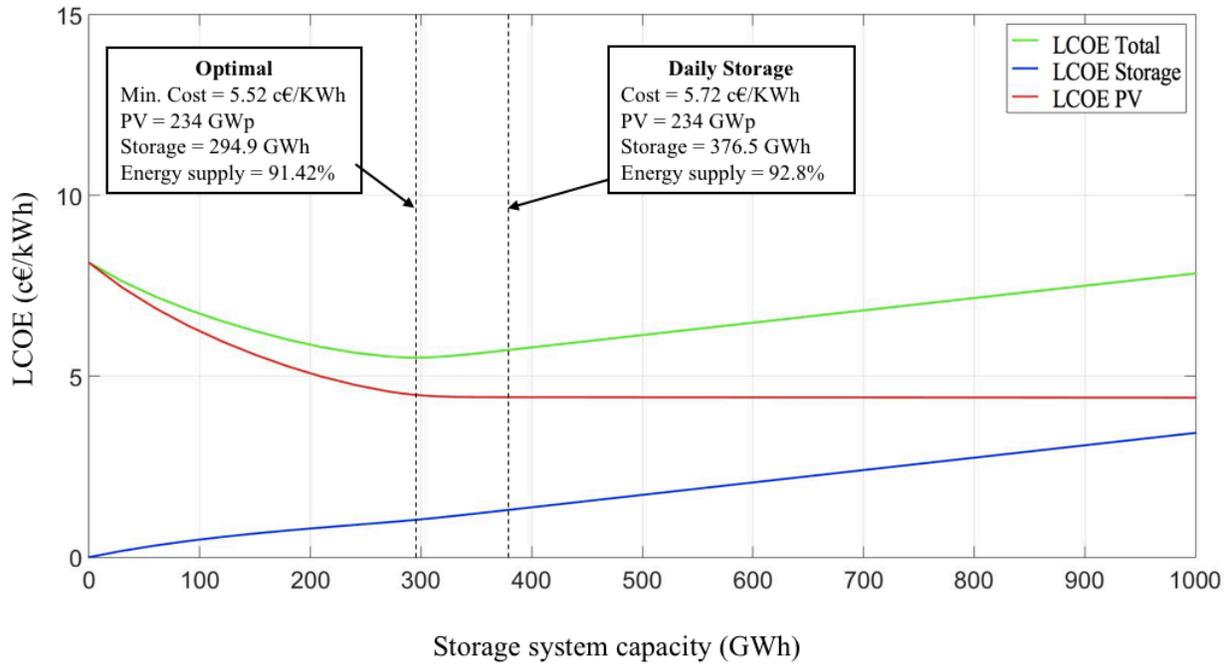

***Supplementary figure 5.3:*** *LCOE estimation vs. battery capacity storage (base case). Only the area around the minimum LCOE value has been represented to present the minimum cost. The complete figure including even seasonal storage is included in the Supplementary figure 3*